\documentclass[a4paper,11pt]{article}
\pdfoutput=1 

\usepackage{jinstpub} 

\usepackage{subcaption}
\usepackage{xcolor}
\usepackage{siunitx}

\title{Development of a Simulation Framework for Spherical Proportional Counters}

\author[a]{I.~Katsioulas,}
\author[a,b]{P.~Knights,}
\author[a]{J.~Matthews,}
\author[a]{T.~Neep,}
\author[a]{K.~Nikolopoulos,}
\author[a]{R.~Owen, }
\author[a,1]{and R.~Ward\note{Corresponding author.}}

\affiliation[a]{School of Physics and Astronomy, University of Birmingham,\\Birmingham, United Kingdom, B15 2TT}
\affiliation[b]{IRFU, CEA, Universite Paris-Saclay,\\ F-91191 Gif-sur-Yvette, France}

\emailAdd{rjw439@bham.ac.uk}

\abstract{The spherical proportional counter is a novel gaseous detector with numerous applications, including direct dark matter searches and neutron spectroscopy. 
  The strengths of the Geant4 and Garfield++ toolkits are combined to create a simulation framework for spherical proportional counters.
  The interface is implemented by introducing Garfield++ classes within a Geant4 application.
  Simulated muon, electron, and photon signals are presented, and the effects of gas mixture composition and anode support structure on detector response are discussed.
}


\arxivnumber{2002.02718} 

\proceeding{15$^{\text{th}}$ Topical Seminar on Innovative Particle and Radiation Detectors\\
  14-17 October 2019\\
  Siena, Italy}

\begin{document}
\maketitle

\section{Introduction}
The spherical proportional counter~\cite{spcInitial} is a novel gaseous detector with a wide range of applications.
These include rare event searches, such as direct low-mass dark matter detection~\cite{Arnaud:2017bjh} and neutrinoless double beta-decay~\cite{Meregaglia:2017nhx};
several forms of spectroscopy, including neutron, alpha, and gamma-ray~\cite{Bougamont:2015jzx};
and low energy neutrino physics~\cite{GIOMATARIS2004330,GIOMATARIS200623,PhysRevD.79.113001}.
A recent review on developments in spherical proportional counter instrumentation is provided in ref.~\cite{spcRecent}.

\begin{figure}[htpb]
\centering
\includegraphics[height = 20em]{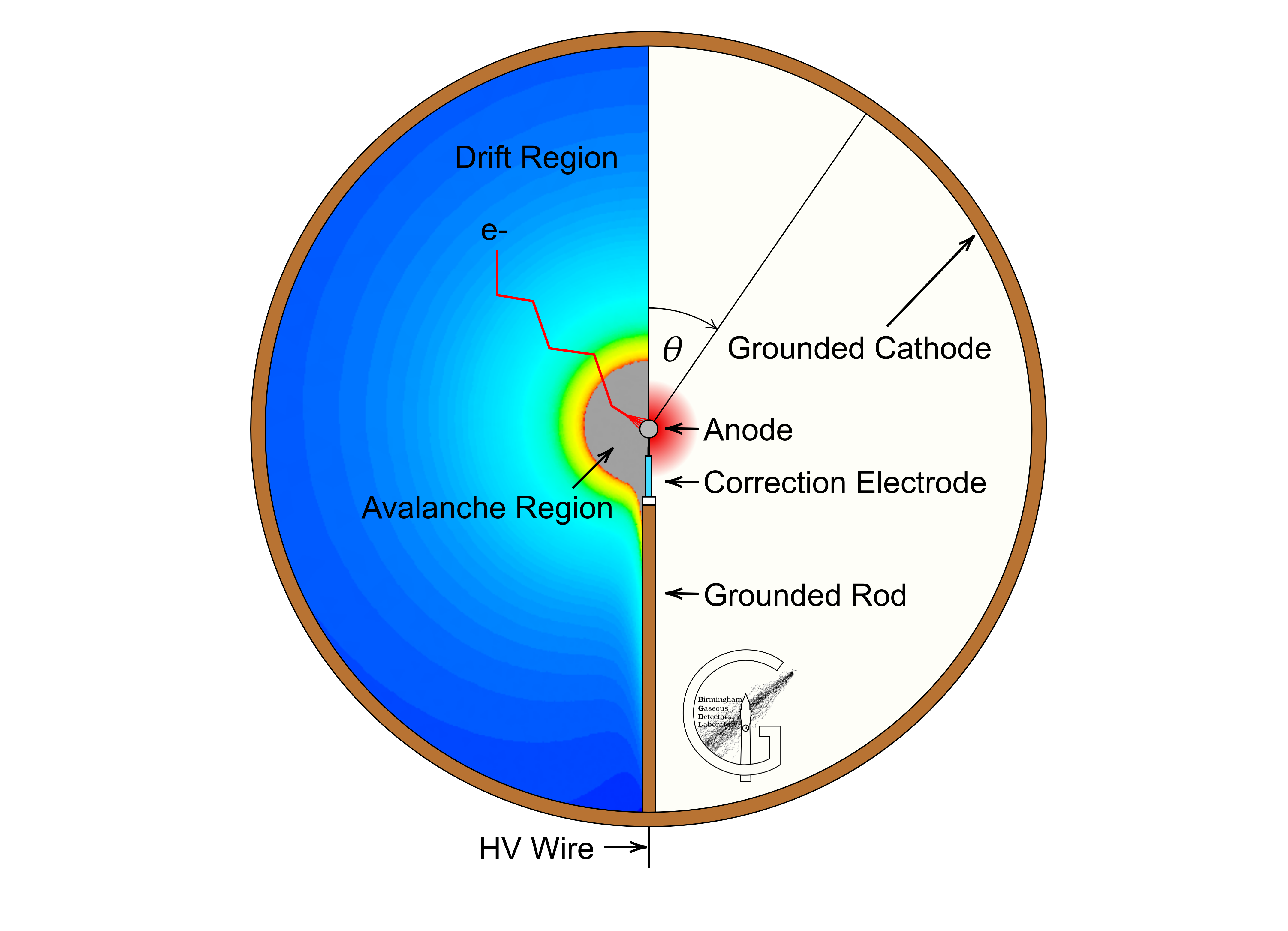}
\caption{Schematic diagram and principle of operation of a spherical proportional counter, including an electric field map calculated with ANSYS finite element method software~\cite{ansys}. $\cos{\theta}~=~1$ corresponds to the top of the detector; $\cos{\theta}~=~-1$ is aligned with the anode support structure.}
\label{figure:schematic}
\end{figure}

\begin{figure}[htpb]
  \centering
  \begin{subfigure}[t]{0.4\textwidth}
    \centering
    \includegraphics[height = 15em]{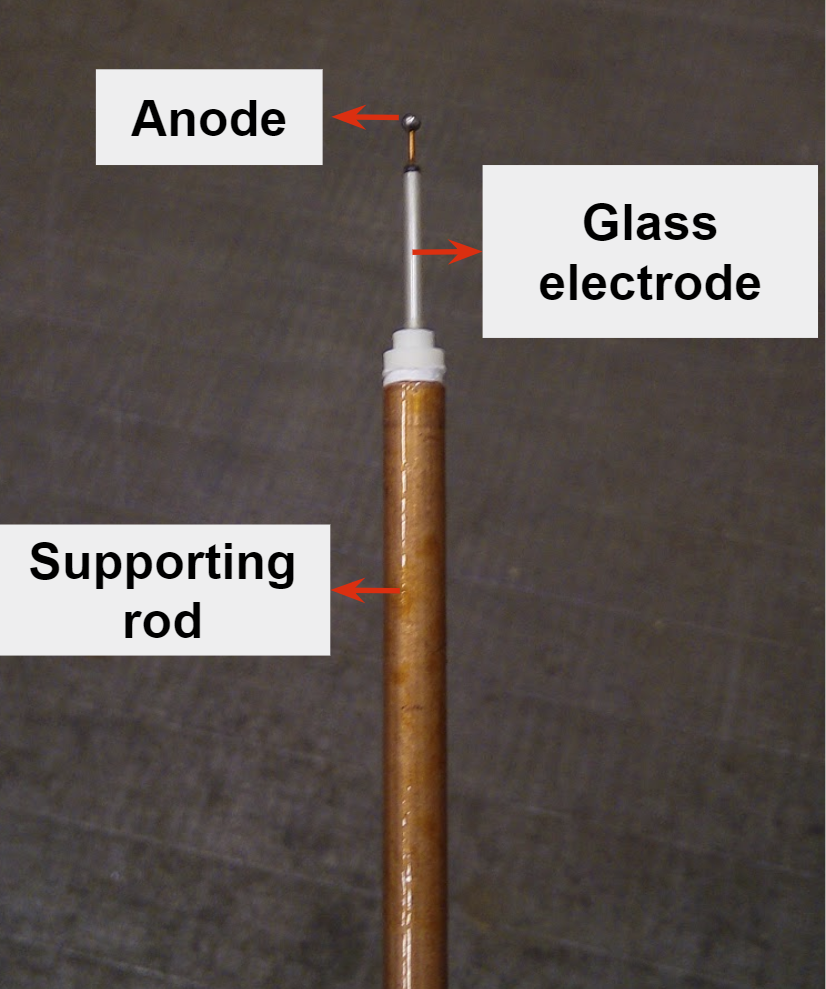} 
    \caption{}
    \label{figure:sensorImage}
  \end{subfigure}
  \begin{subfigure}[t]{0.4\textwidth}
    \centering
    \includegraphics[height = 15em]{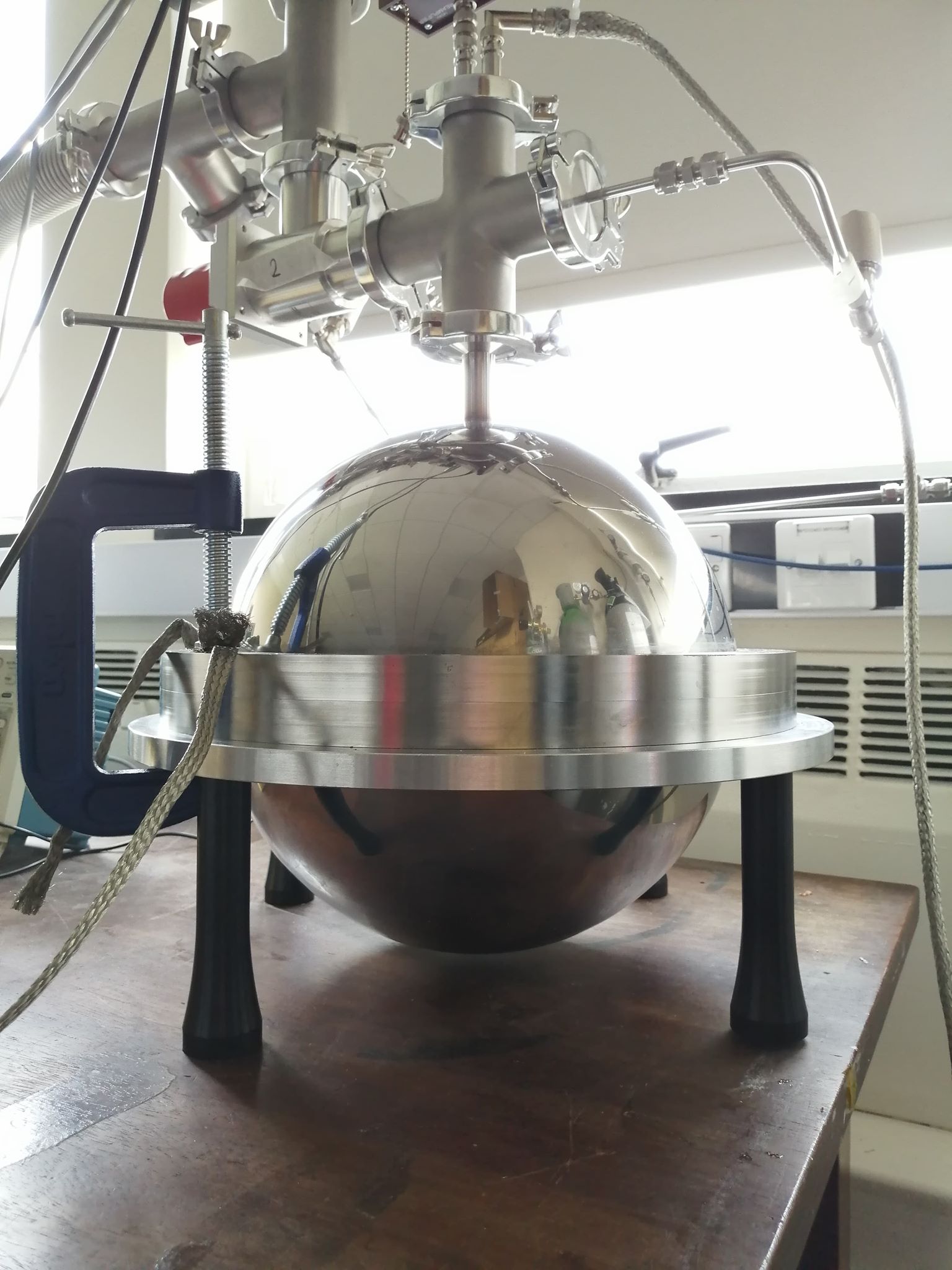}
    \caption{}
    \label{figure:detectorImage}
  \end{subfigure}
  \caption{(\subref{figure:sensorImage}) A sensor with its support structure~\cite{sensors}. (\subref{figure:detectorImage}) A $15$\;cm in radius spherical proportional counter at the University of Birmingham.}
  \label{figure:images}
\end{figure}

A spherical proportional counter is presented schematically in figure~\ref{figure:schematic}. 
It comprises a spherical grounded shell acting as the cathode and enclosing a gas volume, and a spherical anode at the centre supported by a grounded rod.
Figure~\ref{figure:sensorImage} shows an example anode support structure.
Voltage is applied on the anode via a wire fed through the rod. 
Figure~\ref{figure:detectorImage} shows a $15$\;cm in radius spherical proportional counter operating at the University of Birmingham Gaseous Detector Laboratory.
The ideal electric field of a spherical proportional counter is radial and scales as the inverse square of the radial distance, $r$, from the detector centre.
However, the presence of the rod distorts the electric field as shown in figure~\ref{figure:schematic}.
Optimisation of the sensor configuration, for instance by using a correction electrode as in figure~\ref{figure:sensorImage}, is a focus of the detector development programme~\cite{sensors, Giganon:2017isb}. 

Interactions of particles in the detector gas lead to primary ionisation.
These primary ionisation electrons drift towards the anode where the high electric field results in electron multiplication.
The electrons and ions produced in pairs during multiplication drift towards the anode and the cathode, respectively, inducing a signal.
The detector response depends on the interacting particle species, the location of the primary ionisation, the gas mixture, and the configuration of the anode.

Simulations are crucial for detector development and deployment in experiments.
A number of software packages exist for detector simulation, each focusing on different aspects.
Geant4 is a toolkit for the simulation of the passage of particles through matter~\cite{Allison:2016lfl}; Garfield++ is a toolkit for the simulation of gaseous particle detectors~\cite{Veenhof:1998tt, garfield} and interfaces to Heed for particle interactions~\cite{heed} and Magboltz for modelling electron transport parameters in gases~\cite{magboltz}.
The electric field in Garfield++ can be described either analytically or with the use of finite element method software like ANSYS~\cite{ansys}.
A demonstration of the combination of these toolkits was shown in ref.~\cite{combinedSimulation}.
Here, the current status of the development of a predictive framework for the simulation of spherical proportional counters is presented.
Currently, the framework is equipped with all the elements required to provide physics output.
As next steps, the simulation will be validated with data and the user interface will be improved to facilitate wider use by the community.

\section{Detector Simulation}

The framework is a Geant4 application that interfaces Garfield++ in two stages: firstly within a custom physics model used for primary ionisation, electron transport and multiplication; and secondly during the formation of the signal.
The Garfield++ physics model is implemented via the physics parameterisation feature of Geant4.
The signal is calculated at the end of each event through the dedicated ``end-of-event-action'' method in Geant4, which is enhanced with functionality from Garfield++.
Following electron multiplication, information is communicated from the custom physics model to the end-of-event-action method for signal formation using a custom hit collection implemented via the sensitive detector feature of Geant4. 

\begin{figure}[htpb]
  \centering
  \includegraphics[height = 25em]{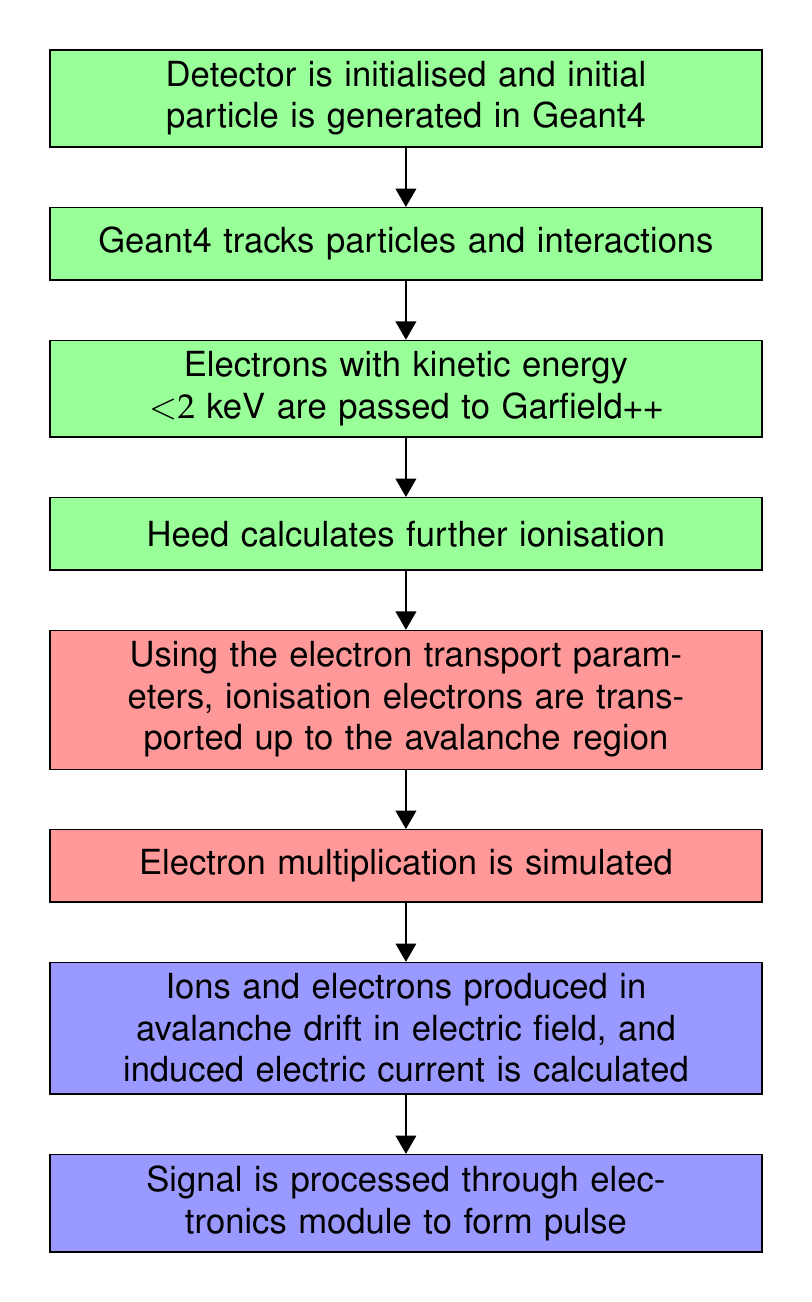}
  \caption{Flow of an event within the simulation framework: primary interactions in green, electron transport and multiplication in red, and signal calculation in blue.}
  \label{figure:workflow}
\end{figure}

The simulation of each event is separated into three main segments, outlined in figure~\ref{figure:workflow}: primary ionisation, electron transport and multiplication within the gas, and signal formation. 

\subsection{Primary Ionisation}
\begin{figure}[htpb]
  \centering
  \includegraphics[height = 15em]{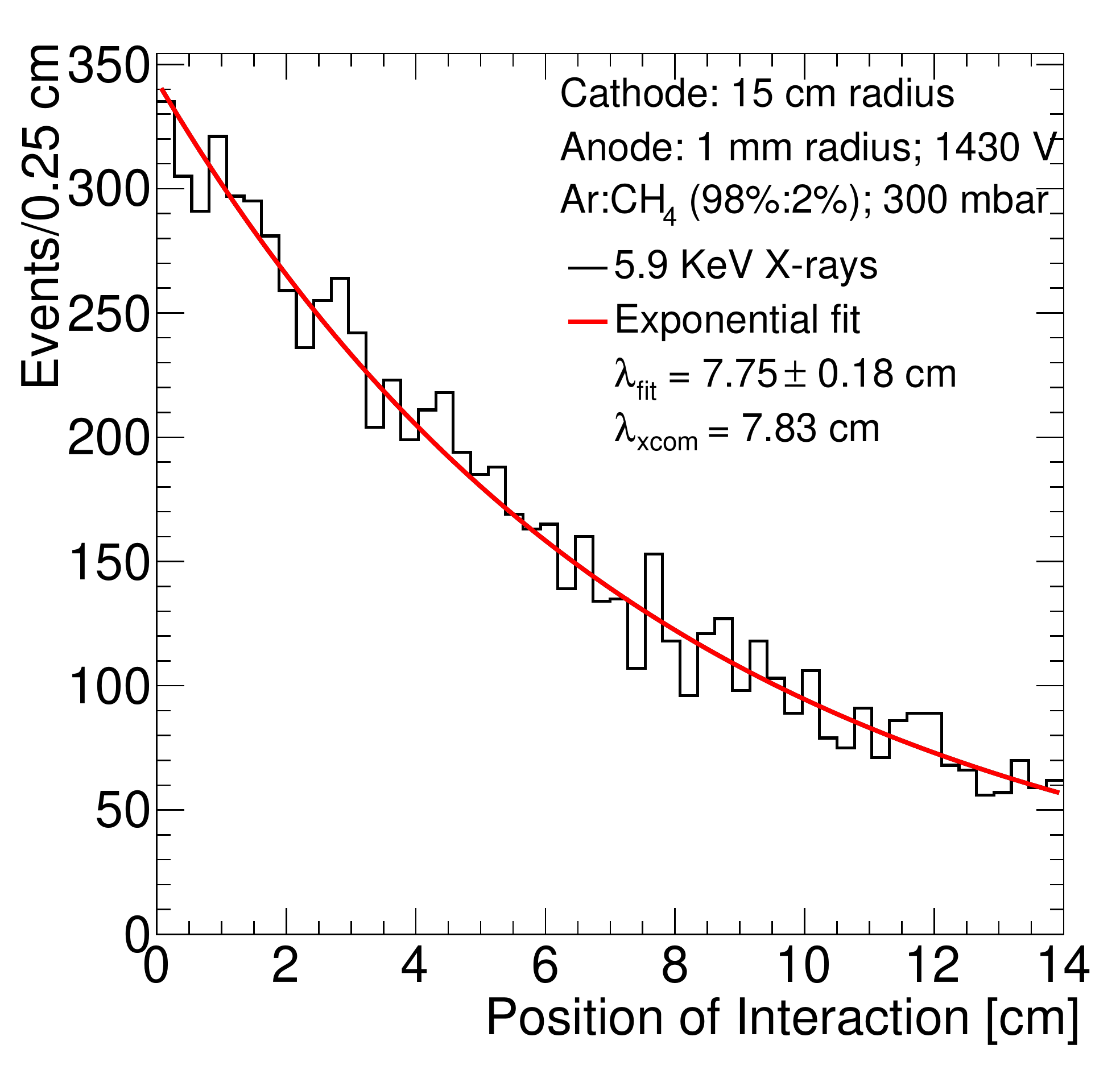}
  \caption{Position of interaction of $5.9$\;keV photons with initial position $r = 14.5$\;cm and $\theta=0$, and direction towards the centre of the detector, including a fit of the decay length compared with the XCOM database~\cite{xcom}.}
  \label{figure:interactionPosition}
\end{figure}

The primary interaction and ionisation by the incident particle is handled in Geant4.
As an example, figure~\ref{figure:interactionPosition} shows the position of interaction for $5.9$\;keV X-rays from the decay of $^{55}$Fe, commonly used for the calibration of detectors, incident towards the centre of the detector from $\theta=0$ and $r = 14.5$\;cm.
The attenuation length modelled by Geant4 is compatible with the XCOM database~\cite{xcom}.

The primary particles ionise the detector gas, producing electrons.
These are modelled with the photo-absorption-ionisation model as implemented in Geant4~\cite{paiModel} until they have a kinetic energy less than $2$\;keV~\cite{combinedSimulation}.
At this threshold, electrons are passed to a custom physics model which uses Heed to calculate the final ionisation, including $\delta$-electron production.

\subsection{Electron Transport and Multiplication}

\begin{figure}[ht]
  \centering
  \begin{subfigure}[t]{0.32\textwidth}
    \includegraphics[height = 12em]{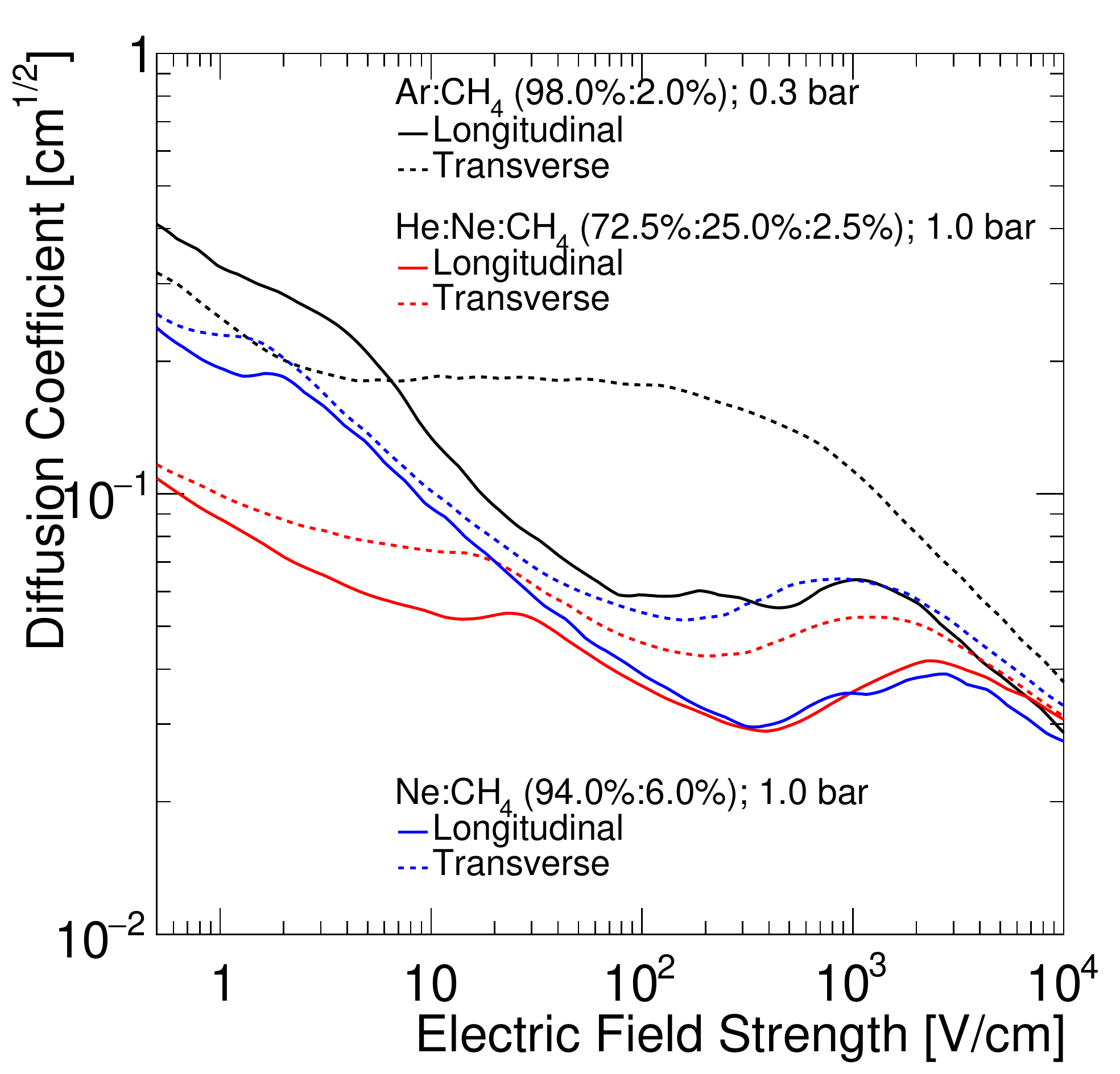}
    \caption{\label{figure:diffusion}}
  \end{subfigure}
  \begin{subfigure}[t]{0.32\textwidth}
    \includegraphics[height = 12em]{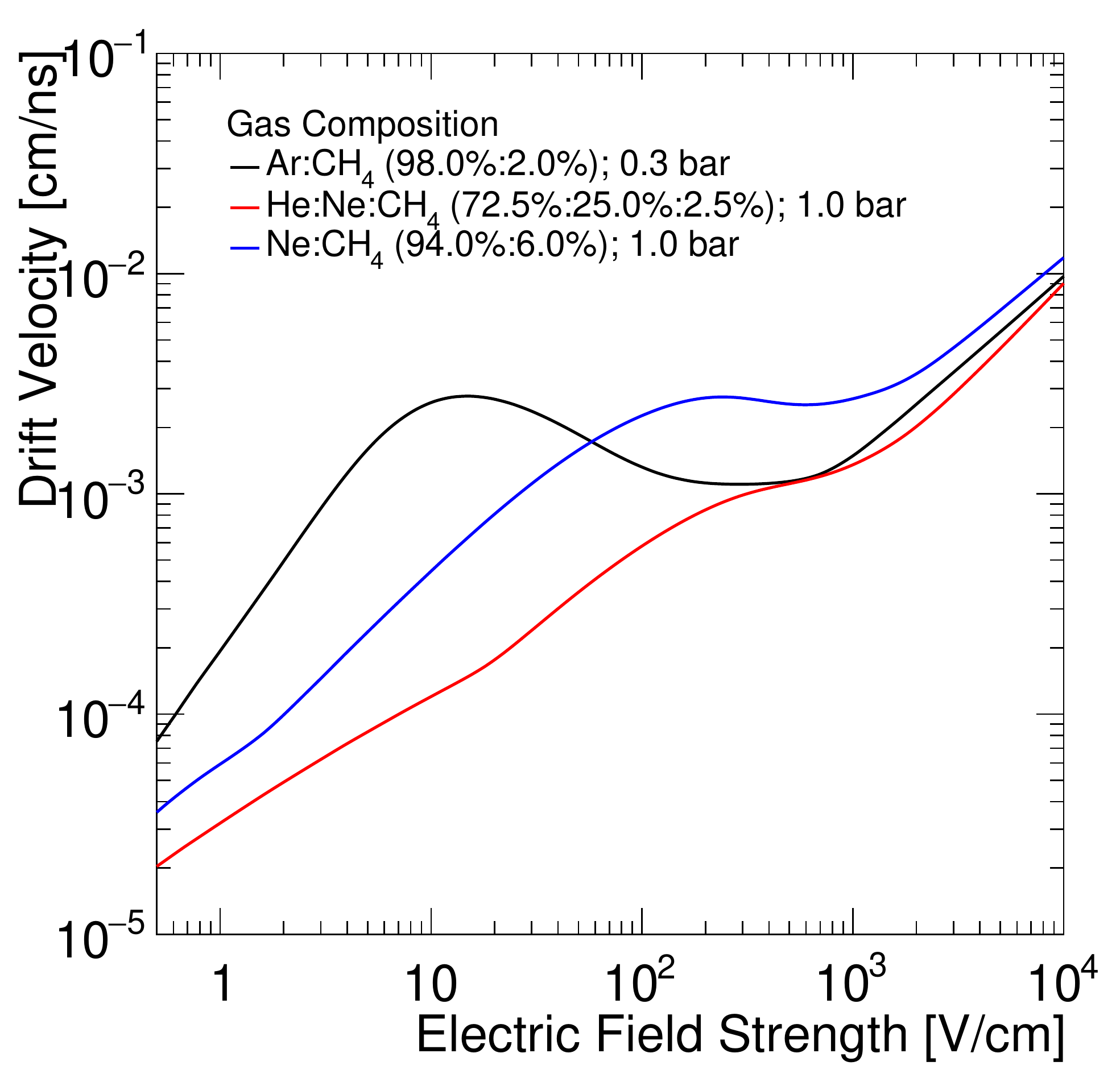}
    \caption{\label{figure:driftv}}
  \end{subfigure}
  \begin{subfigure}[t]{0.32\textwidth}
    \includegraphics[height = 12em]{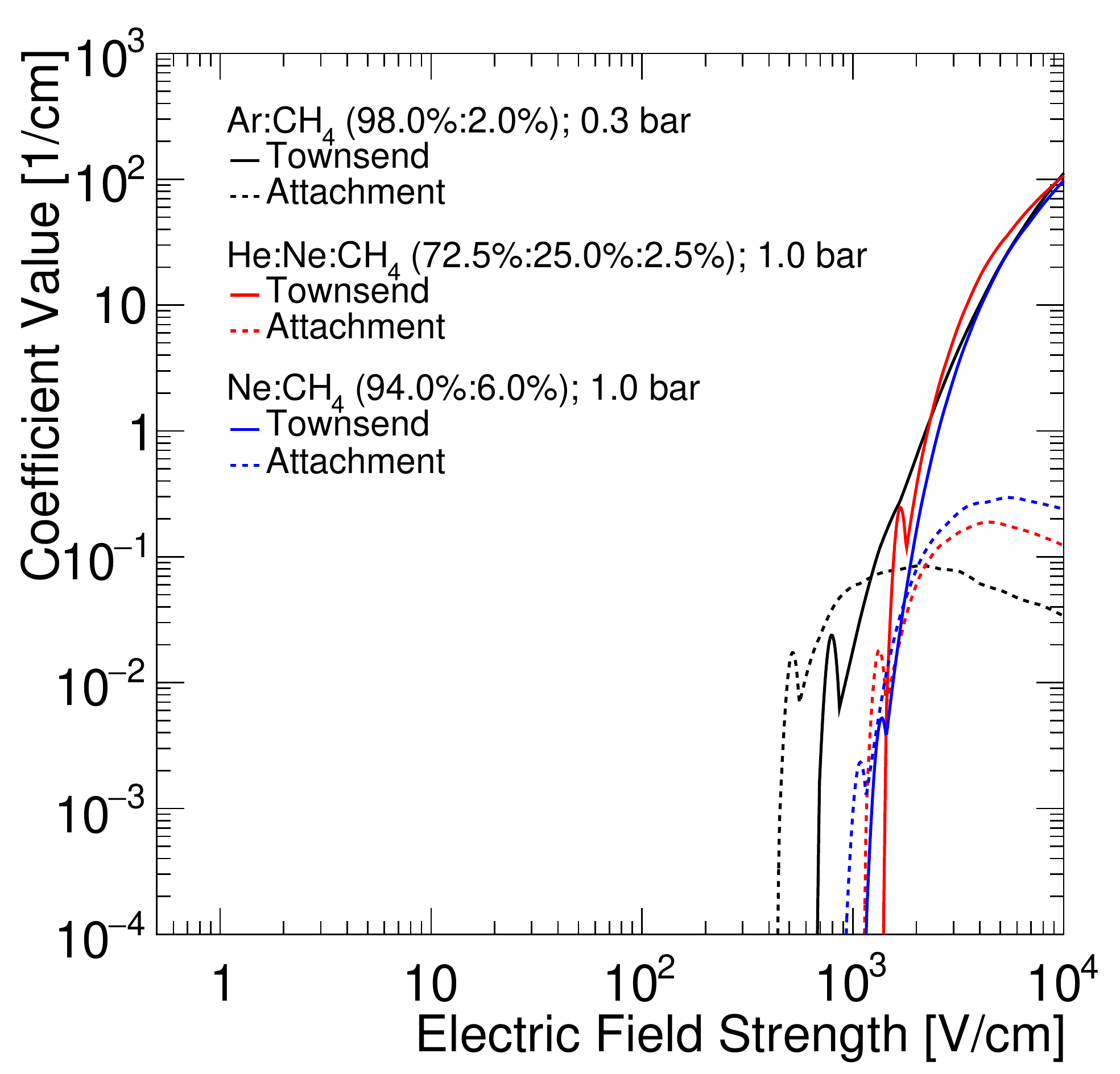}
    \caption{\label{figure:avalancheCoefficients}}
  \end{subfigure}
  \caption{Electron transport parameters versus the electric field strength for several gas mixtures, calculated by Magboltz: (\subref{figure:diffusion})~longitudinal and transverse diffusion coefficients, (\subref{figure:driftv})~drift velocity, and (\subref{figure:avalancheCoefficients})~Townsend and attachment coefficients.}
  \label{figure:magboltz}
\end{figure}

The ionisation electrons are transported through the gas stochastically.
For this purpose the Monte Carlo drift line method of Garfield++ is used.
Electrons are transported in discrete steps; the new electron position at each step is obtained by integrating the electron's equation of motion in the electric field, and subsequently adding a randomly sampled diffusion step.
Magboltz gas property tables are used to model the electron drift velocity and diffusion in the electric field, as well as the Townsend, $\alpha$, and attachment, $\eta$, coefficients.
Figures~\ref{figure:diffusion} and~\ref{figure:driftv} show electron longitudinal and transverse diffusion coefficients, and drift velocities, respectively, for the gases used in this work.
Figure~\ref{figure:avalancheCoefficients} shows the corresponding electron Townsend and attachment coefficients.

When the electrons approach the anode two options are provided in Garfield++ to model the multiplication: microscopic drift line tracking and Monte Carlo drift line tracking.
The microscopic drift line tracking models the multiplication process down to the individual electron-atom collision level.
As such, it is the most precise option, but it has a large computational cost.
For this reason, in this work an additional customised avalanche option was developed.

\begin{figure}[htpb]
  \centering
    \includegraphics[height = 15em]{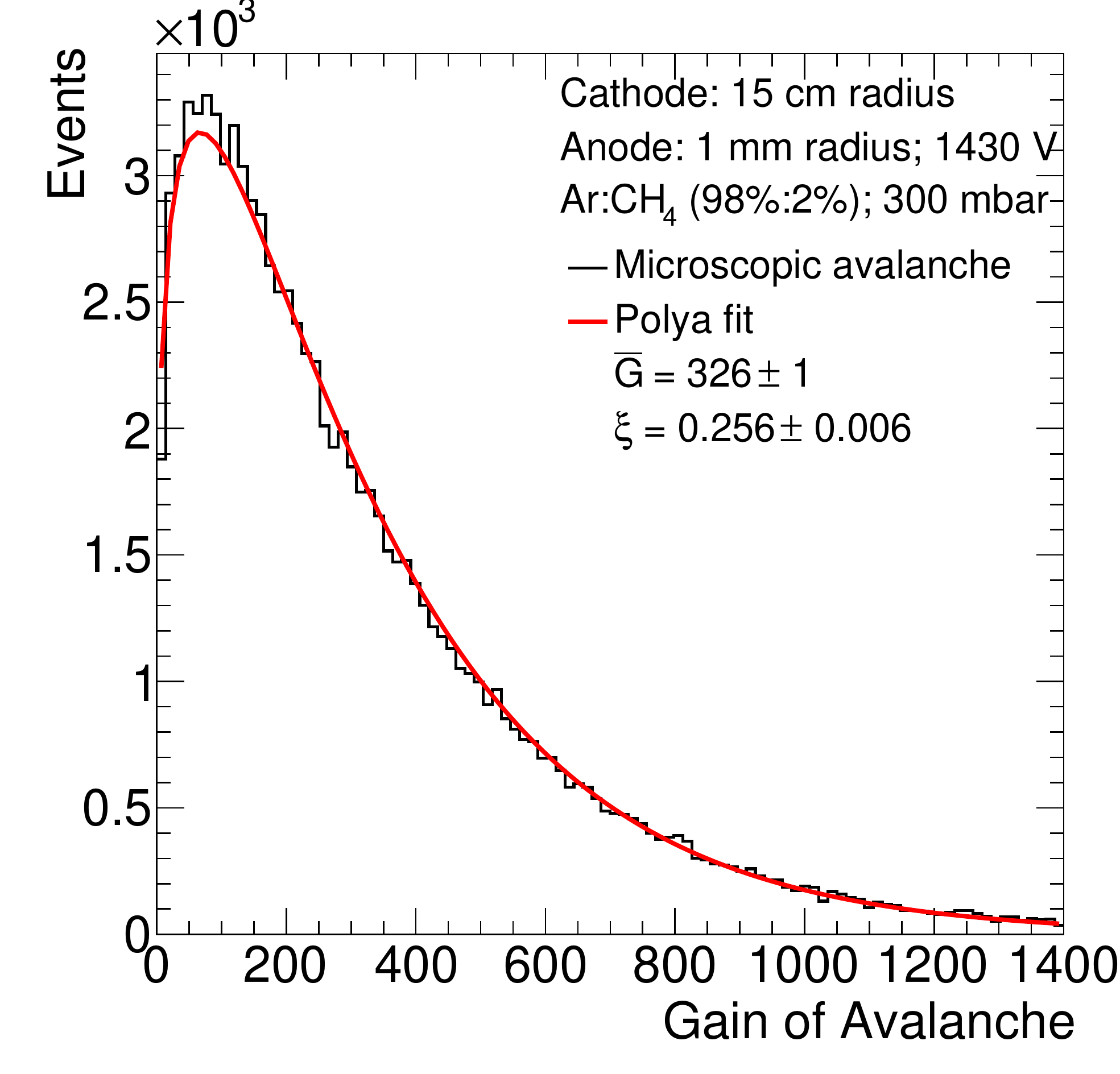}
  \caption{Simulation of the electron avalanche. Gain calculated with Garfield++ microscopic avalanche tracking (black) and fit with a Polya distribution (red).}
  \label{figure:avalanche}
\end{figure}

In the custom multiplication model, the average gain, $\bar{G}$, is calculated as 
\begin{equation}
  \bar{G} = \exp{\bigg[ \int_{\vec{r}}{ \Big(\alpha(\textbf{r}) - \eta(\textbf{r})\Big)d\vec{r}} \bigg] }.
  \label{equation:averageGain}
\end{equation}
The integral is evaluated numerically as the electron approaches the anode.
Fluctuations with respect to the average gain, $G/\bar{G}$, are modelled using the Polya distribution~\cite{knoll}, 
\begin{equation}
  P\bigg(\frac{G}{\bar{G}}\bigg) = \bigg((1 + \xi)\frac{G}{\bar{G}}\bigg)^\xi \exp{\bigg[-(1 + \xi)\frac{G}{\bar{G}}\bigg]},
  \label{equation:polya}
\end{equation}
which has a width parameter, $\xi$.
Parameter $\xi$ is estimated by running a simulation with Garfield++ microscopic tracking and fitting the result, as in figure~\ref{figure:avalanche}; $\xi$ was found to be approximately independent with respect to the position of the avalanche.

\subsection{Signal Formation}

Following the avalanche creation, for each ionisation electron that undergoes multiplication a single ``electron'' and a single ``ion'' with charges equal to $-(G + 1)|e|$ and $G|e|$, respectively, are created.
It is found that this approximation significantly improves the code performance in terms of CPU, without appreciable loss of information. 
These ``electron-ion'' pairs are passed to the Geant4 end-of-event action for signal formation, where the current signal is estimated using Garfield++ to simulate the drift of the ions and electrons.
Garfield++ calculates the current induced by each ``electron-ion'' pair using the Shockley-Ramo theorem~\cite{shockley, ramo}. 

\begin{figure}[htpb]
  \centering
  \begin{subfigure}[t]{0.4\textwidth}
    \centering
    \includegraphics[height = 15em]{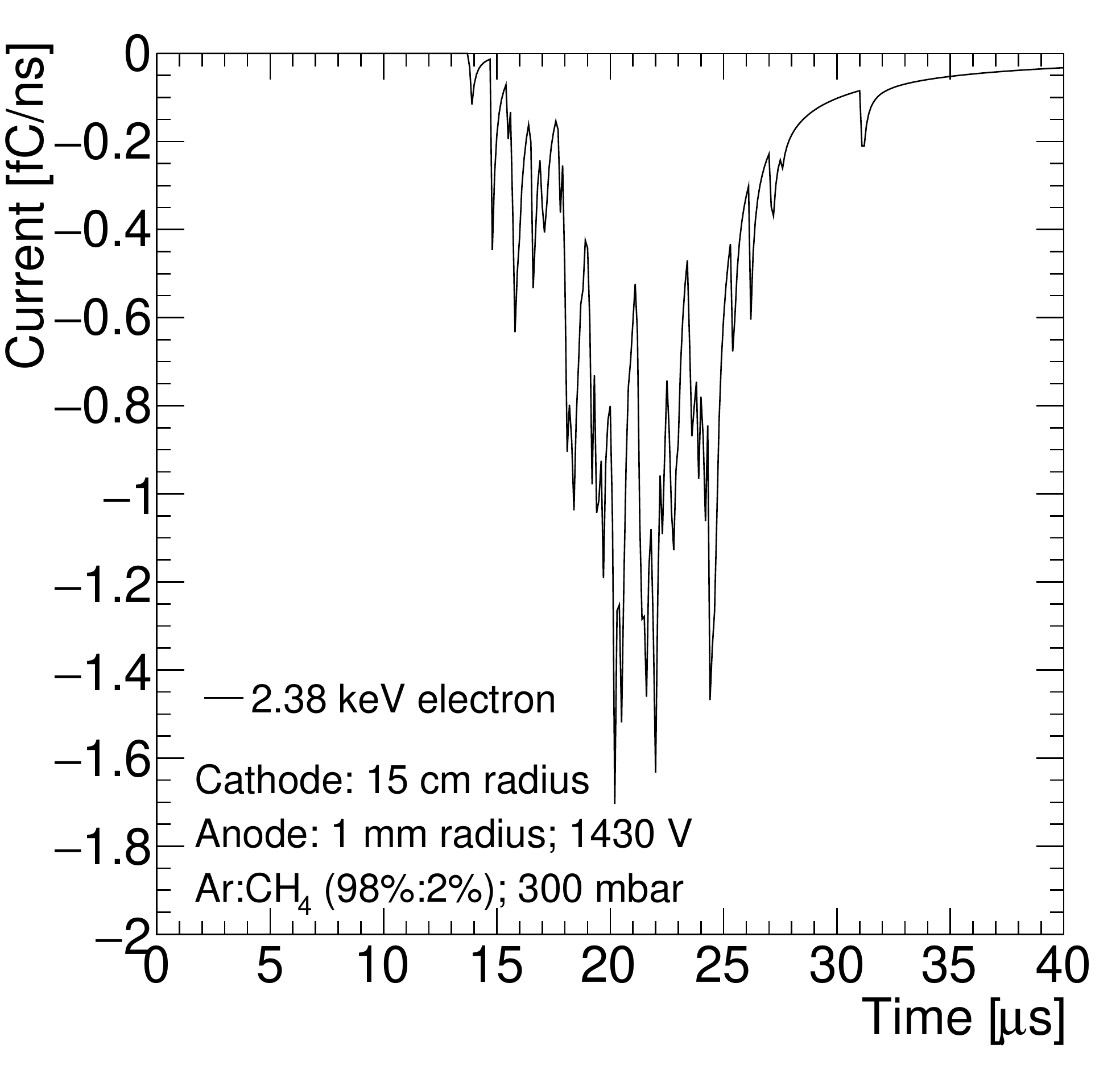}
    \caption{Current signal}
    \label{figure:signalCurrent}
  \end{subfigure}
  \begin{subfigure}[t]{0.4\textwidth}
    \centering
    \includegraphics[height = 15em]{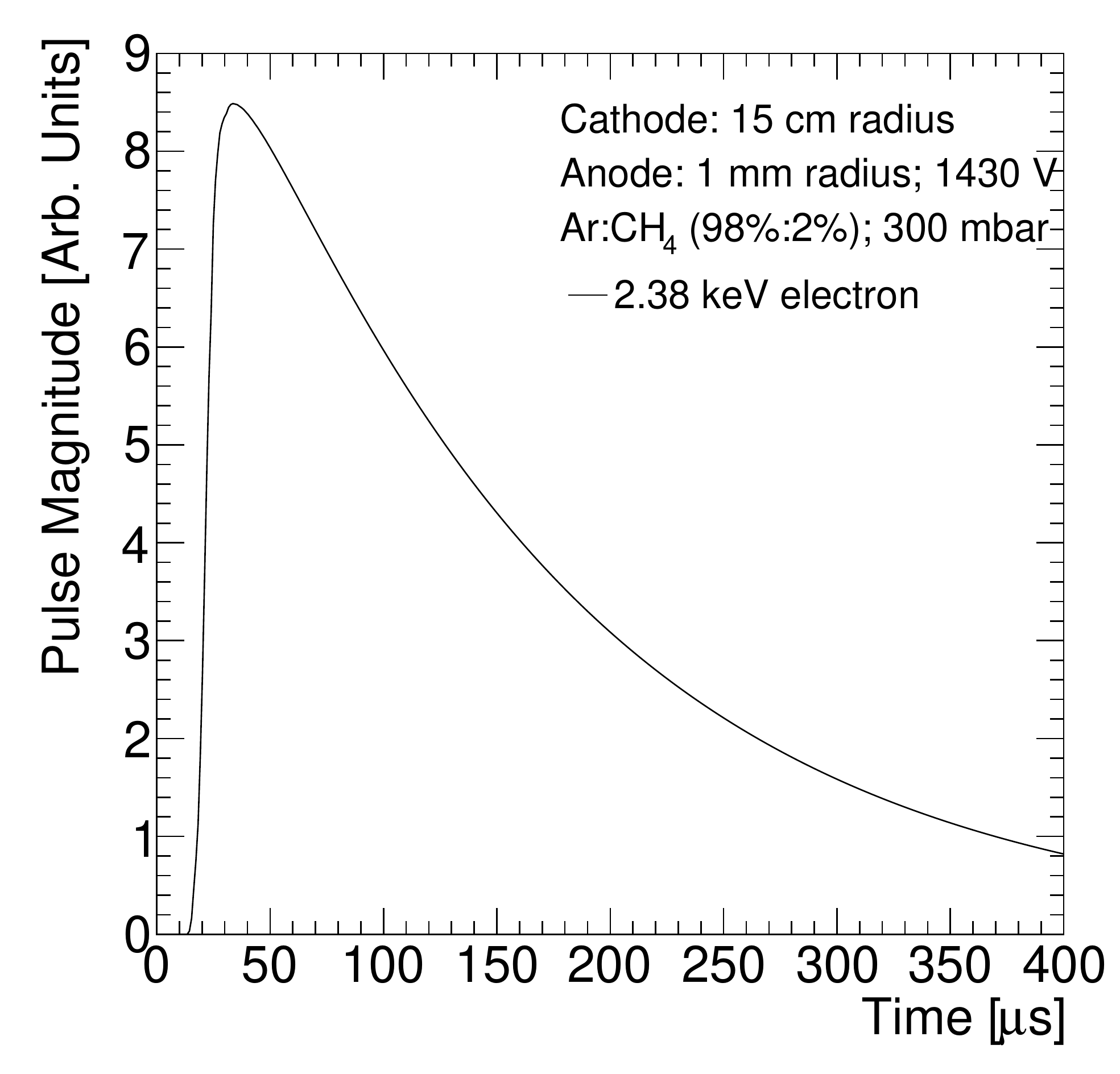}
    \caption{Voltage signal}
    \label{figure:signalVoltage}
  \end{subfigure}
  \caption{(\subref{figure:signalCurrent}) The current induced and (\subref{figure:signalVoltage}) the readout pulse produced by a $2.38$\;keV electron interacting in the gas Ar:CH$_{4}~(98\%:2\%)$ at $300$\;mbar from an initial radius of $10$\;cm and $\theta = 0$.}
  \label{figure:signal}
\end{figure}

An example simulated current signal produced by the interaction of a $2.38$\;keV Auger electron, from the decay of $^{37}$Ar to $^{37}$Cl via electron capture~\cite{nndc:ar37}, is shown in figure~\ref{figure:signalCurrent}.
The arrival of each ionisation electron at the anode results in distinct spikes in the current as each avalanche occurs.
The current signal is integrated and processed through an electronics module to form the voltage pulse, as shown in figure~\ref{figure:signalVoltage}.
In this case, the transfer function of a simple charge sensitive amplifier with an integration time constant of $140\;$\si{\micro\second} is used.

\section{Simulation Results}
A $15$\;cm in radius detector was simulated using a $1$\;mm radius anode at $1430$\;V and several gas mixtures.
Two configurations were investigated: an ideal detector with the analytic field and a realistic sensor, similar to that in figure~\ref{figure:sensorImage} with an electric field calculated with ANSYS.

\subsection{Effect of the Gas Mixture Composition}

\begin{figure}[htpb]
  \centering
  \includegraphics[height = 15em]{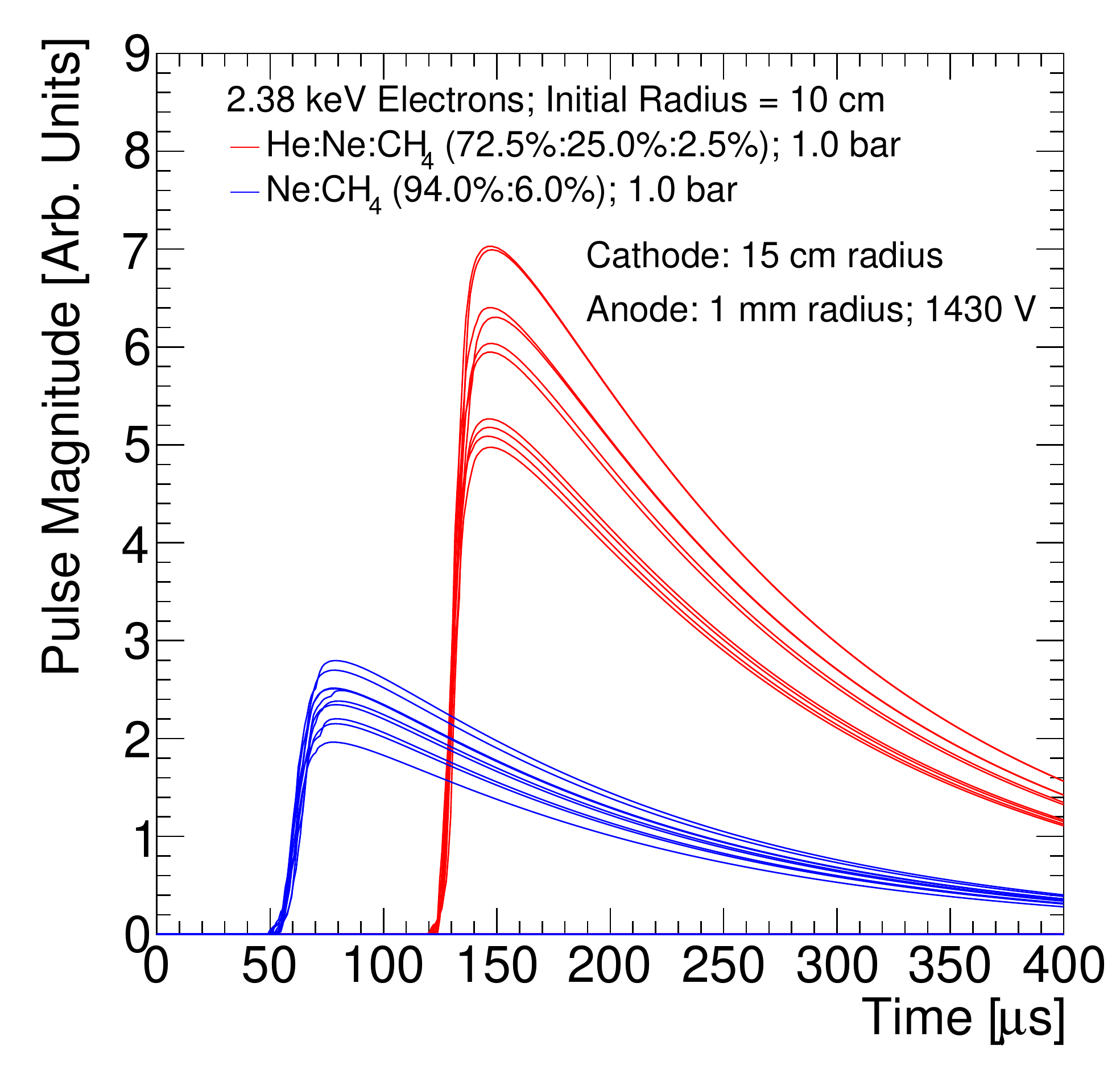}
  \caption{Readout pulses produced by $2.38$\;keV electrons from an initial radius of $10$\;cm in a $15$\;cm in radius detector in the gases He:Ne:CH$_{4}~(72.5\%:25.0\%:2.5\%)$ at $1.0$\;bar (red) and Ne:CH$_{4}~(94\%:6\%)$ at $1.0$\;bar (blue).}
  \label{figure:pulse}
\end{figure}

Figure~\ref{figure:pulse} shows example pulses produced by $2.38$\;keV electrons, with identical initial positions, in two different gases with the realistic configuration, demonstrating a number of features.
The mean amplitude of the pulses in He:Ne:CH$_{4}~(72.5\%:25.0\%:2.5\%)$ is approximately $2.5$ times larger than that in Ne:CH$_{4}~(94\%:6\%)$, as expected from the Townsend and attachment coefficients shown in figure~\ref{figure:avalancheCoefficients}.
The gain fluctuations are demonstrated by the variance in amplitude between pulses produced under the same conditions. The time at which the pulses start forming is different, as expected by the difference in the drift velocity curves in figure~\ref{figure:driftv}, where He:Ne:CH$_{4}~(72.5\%~:~25.0\%~:~2.5\%)$ has a smaller drift velocity thoughout the electric field range of the detector.

\subsection{Effect of the Anode Support Structure}

\begin{figure}[htpb]
  \centering
  \begin{subfigure}[t]{0.4\textwidth}
    \centering
    \includegraphics[height = 15em]{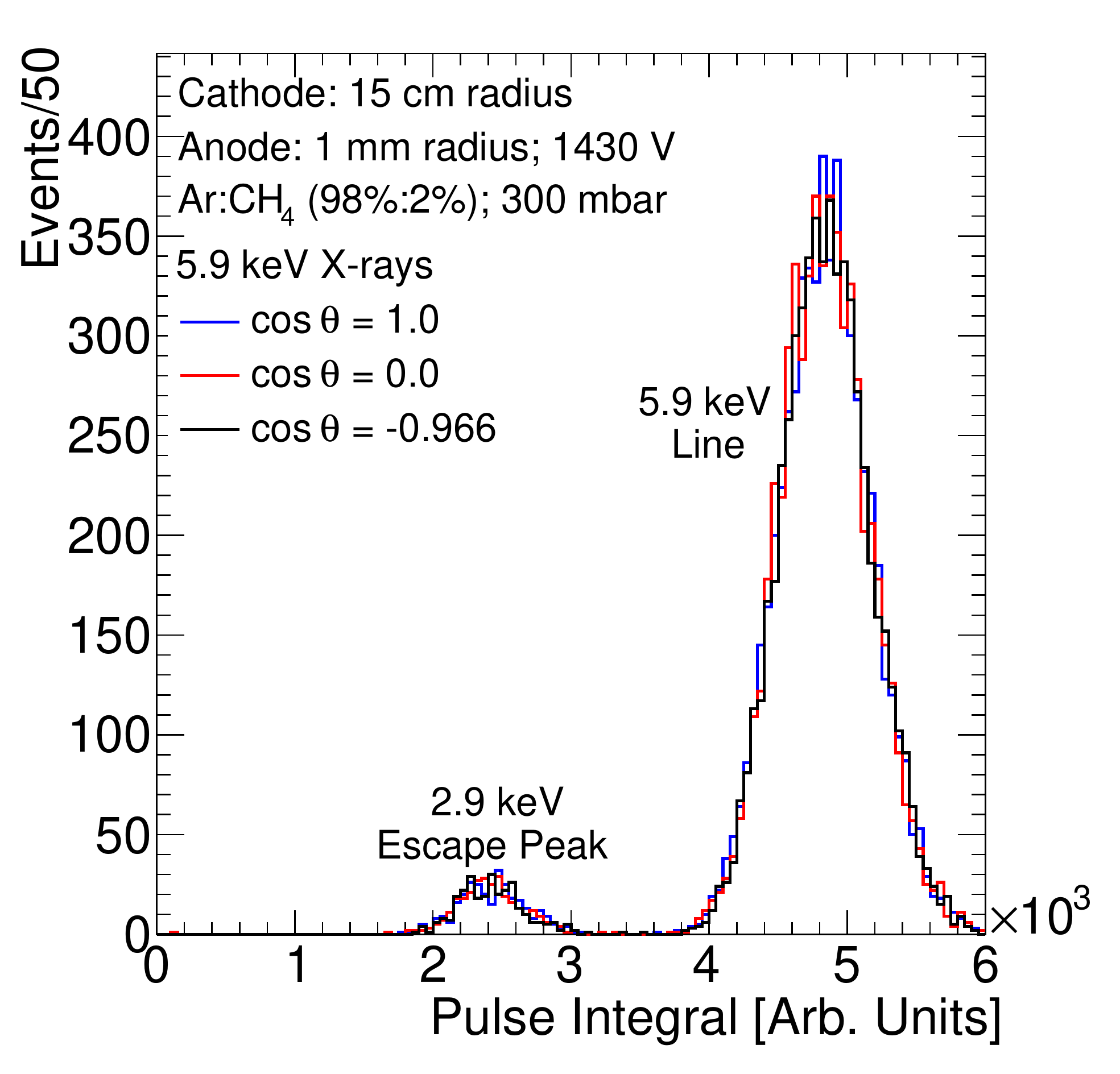}
    \caption{Ideal detector}
    \label{figure:idealIntegral}
  \end{subfigure}
  \begin{subfigure}[t]{0.4\textwidth}
    \centering
    \includegraphics[height = 15em]{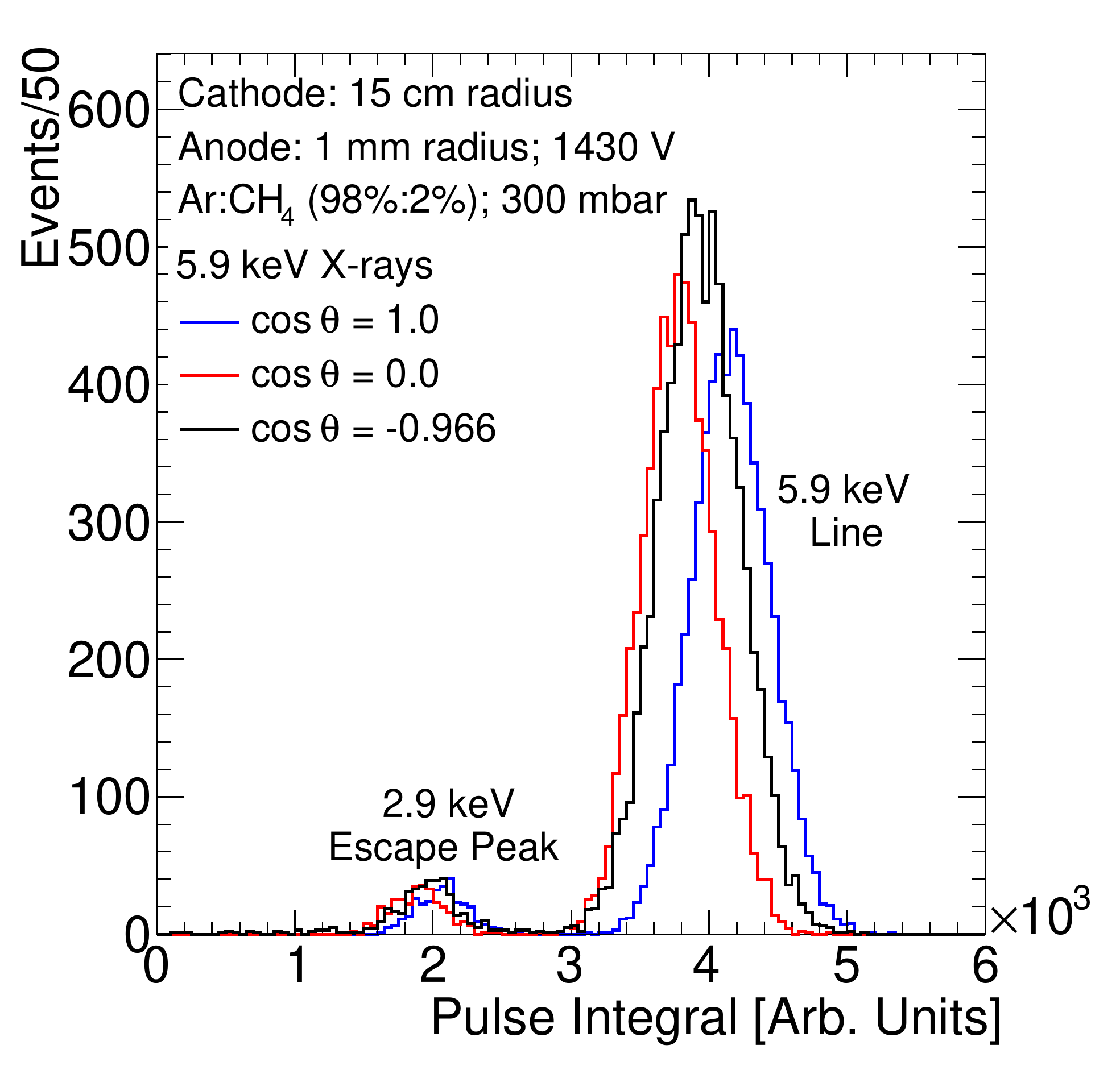}
    \caption{Realistic configuration}
    \label{figure:umbrellaIntegral}
  \end{subfigure}
  \caption{The pulse integral from interactions of $5.9$\;keV electrons in Ar:CH$_{4}~(98\%:2\%)$ at $300$\;mbar in a $15$\;cm in radius detector, (\subref{figure:idealIntegral}) using an ideal detector; (\subref{figure:umbrellaIntegral}) using a realistic configuration with a correction electrode and a supporting rod.}
  \label{figure:pulseIntegral}
\end{figure}

The simulation framework can be used to investigate how the anode support structure affects detector response.
Figure~\ref{figure:idealIntegral} shows the pulse integral analysis of $5.9$\;keV photon signals in an ideal detector, incident from several angles, in Ar:CH$_{4}~(98\%:2\%)$ gas at $300$\;mbar.
Two distinct peaks are measured: the $5.9$\;keV line and the argon escape peak at $2.9$\;keV.
In the ideal detector the signal does not change as a function of $\theta$.
Conversely, with the realistic configuration $\theta$ does affect the response, as shown in figure~\ref{figure:umbrellaIntegral}. 

\subsection{Interaction Radius and Detector Fiducialisation}

\begin{figure}[htpb]
  \centering
  \begin{subfigure}[t]{0.32\textwidth}
    \centering
    \includegraphics[height = 12em]{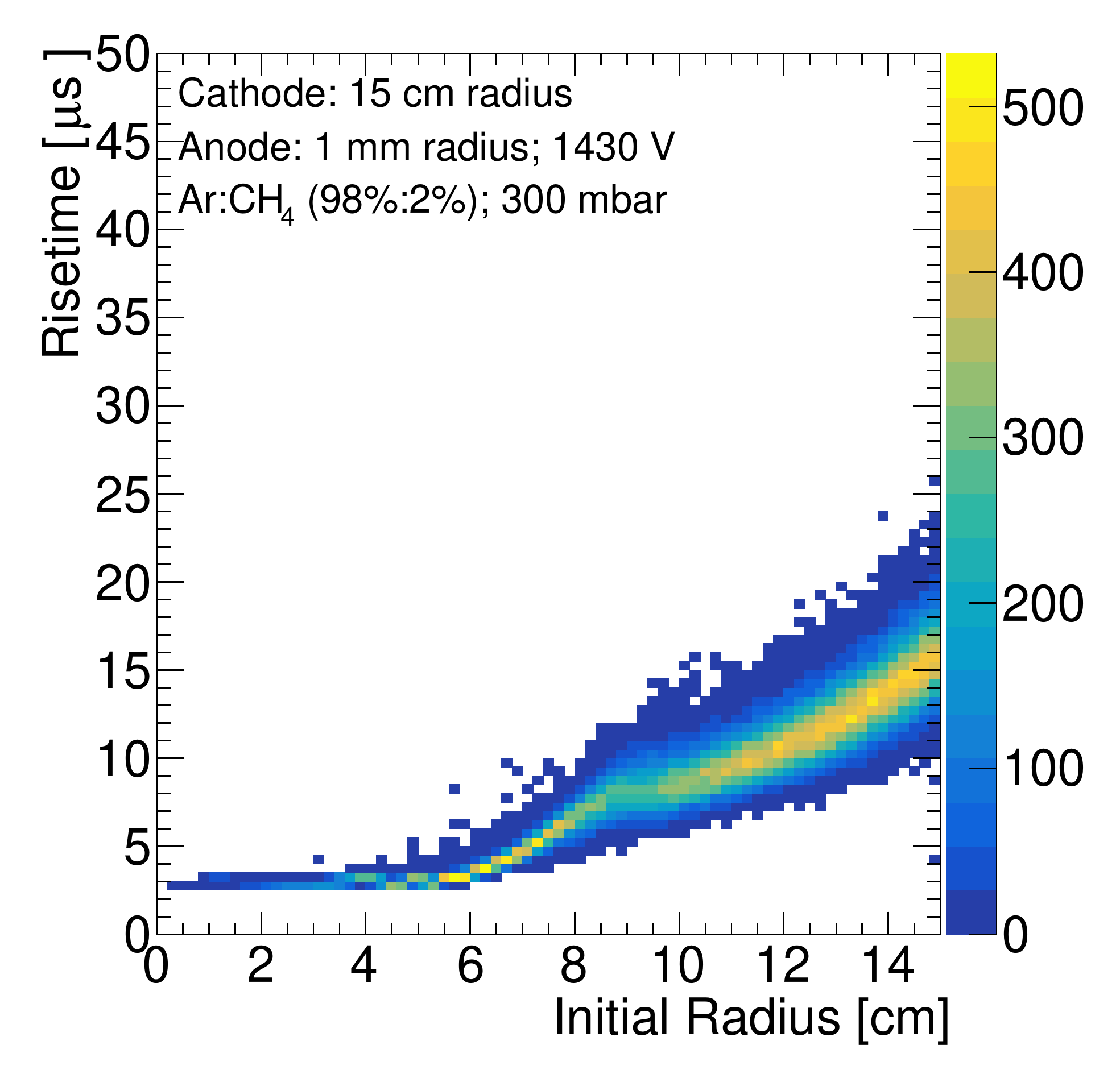}
    \caption{\label{figure:idealTotal}}
  \end{subfigure}
  \begin{subfigure}[t]{0.32\textwidth}
    \centering
    \includegraphics[height = 12em]{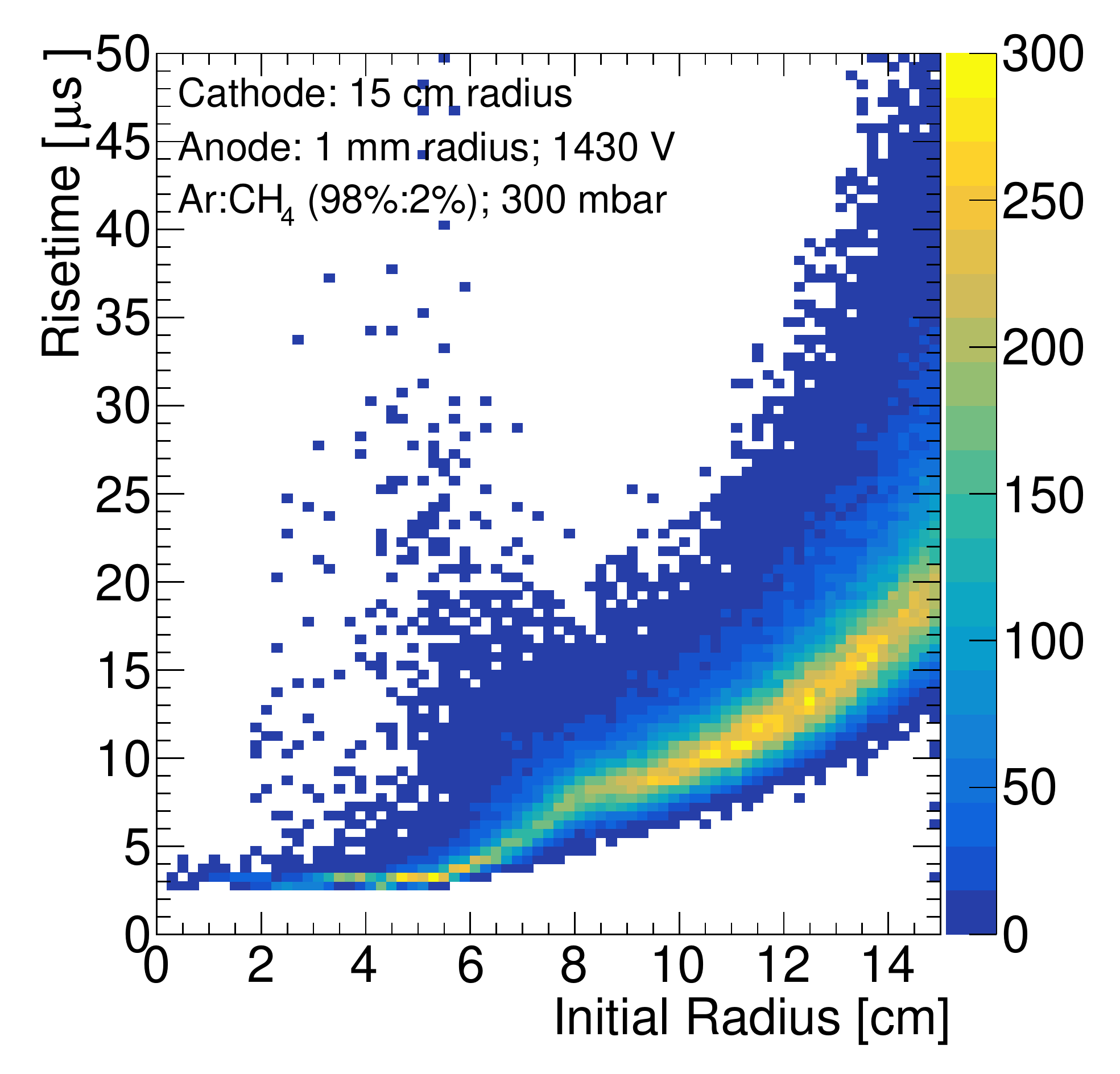}
    \caption{\label{figure:umbrellaTotal}}
  \end{subfigure}
  \begin{subfigure}[t]{0.32\textwidth}
    \centering
    \includegraphics[height = 12em]{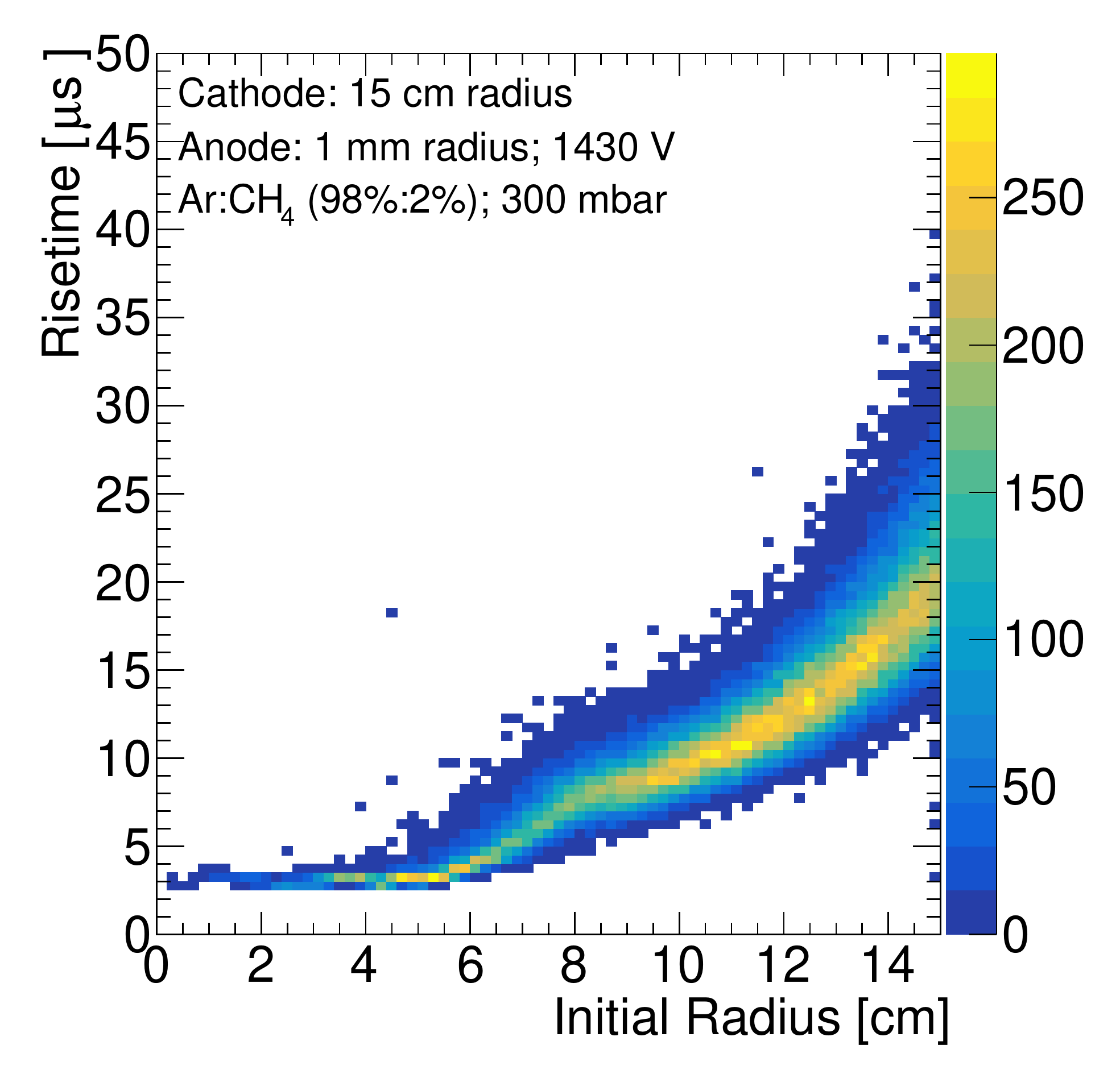}
    \caption{\label{figure:umbrellaCosTheta}}
  \end{subfigure}
  \caption{The increase of pulse risetime with interaction radius with $2.38$\;keV electrons interacting uniformly in a $15$\;cm in radius detector. (\subref{figure:idealTotal}) Interactions in an ideal detector;  (\subref{figure:umbrellaTotal}) interactions using the realistic configuration, inclusive of events near the support structure; (\subref{figure:umbrellaCosTheta}) interactions using a realistic configuration selecting events with $\cos{\theta > - 0.7}$ on the initial electron angular position.}
  \label{figure:interactionRadius}
\end{figure}

Ionisation electrons produced in events at the edge of the detector drift longer than electrons produced near the centre.
As a result of the increased diffusion, the risetime of pulses produced at larger radii increases, where risetime is defined as the time for a pulse to rise from $10\%$ to $90\%$ of its amplitude.
Figure~\ref{figure:interactionRadius} demonstates this with the interaction of $2.38$\;keV electrons uniformly in the detector, using a $300$\;mbar Ar:CH$_{4}~(98\%:2\%)$ gas mixture.
These electrons have a short range, $1.3$\;mm~\cite{estar}, and deposit their energy near their initial position.
In rare event searches, a substantial fraction of background events originate from the detector surface; the radial dependence of the risetime allows the detector to be fiducialised. 

Figure~\ref{figure:idealTotal} shows interactions in an ideal detector, while figure~\ref{figure:umbrellaTotal} shows interactions using the realistic configuration. 
The distortion in the electric field due to the presence of the rod increases the length of the path ionisation electrons travel to the anode.
This increases diffusion in events near the rod, leading to a population of events with increased risetimes not present in the ideal case.
Selecting events away from the rod, with initial $\cos{\theta} > -0.7$, removes that population, as shown in figure~\ref{figure:umbrellaCosTheta}.

\subsection{Particle Identification}
\begin{figure}[!htpb]
  \centering
  \includegraphics[height = 15em]{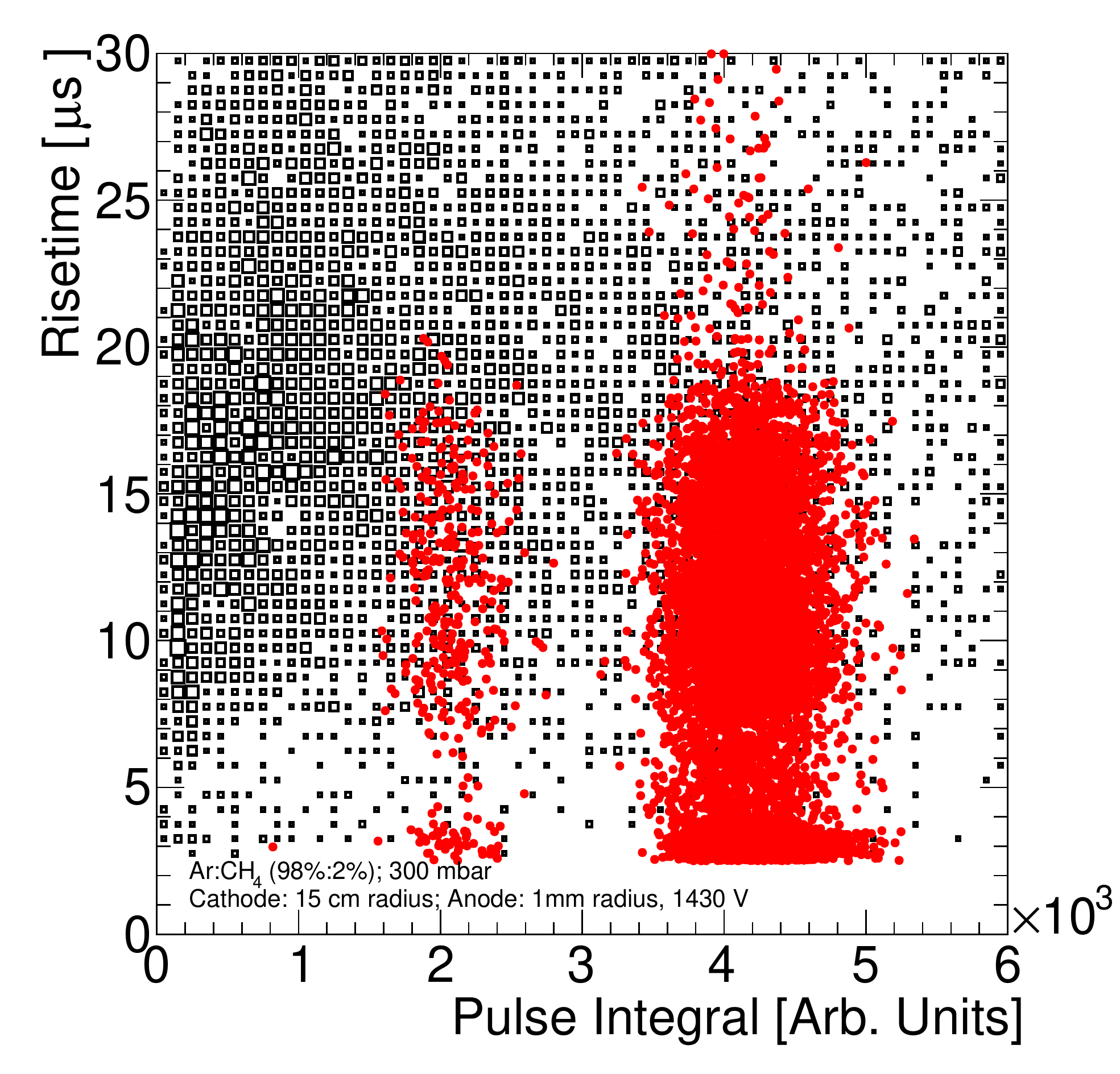}
  \caption{Pulse shape property comparison for signals produced by cosmic-ray muons (black boxes) and $5.9$\;keV X-rays (red points) in Ar:CH$_{4}~(98\%:2\%)$ at $300$\;mbar in a $15$\;cm in radius detector.\label{figure:particleIdentification}}
\end{figure}

Figure~\ref{figure:particleIdentification} compares pulses produced by
cosmic-ray muons and $5.9$\;keV X-rays, using a $300$\;mbar
Ar:CH$_{4}~(98\%:2\%)$ gas mixture.  The black boxes are pulses
produced by cosmic-ray muons, which have a range of initial
energies~\cite{PhysRevD.98.030001} and traverse different paths
through the detector. This leads to significant variations in the
total ionisation in the gas.  Of the red points, the rightmost
population is the $5.9$\;keV line and the leftmost population is the
$2.9$\;keV argon escape peak, as in figure~\ref{figure:pulseIntegral}.
As the ionisation profile for muons and X-rays is different,
pulse-shape analysis can be used for particle identification and
discrimination.

\section{Summary}

A flexible and predictive framework for the simulation of the spherical proportional counter is developed, combining the strengths of the Geant4 and Garfield++ toolkits. 
This framework allows detector response to be studied under different conditions, facilitating investigations of sensor development, and event reconstruction.
Examples of simulated events with different particle species, detector configurations, and gas mixtures were presented and discussed.
In the future, the simulation will be validated with data and the user interface will be improved to facilitate the wider use of this framework by the community.

\acknowledgments 
This project has received funding from the European
Union's Horizon 2020 research and innovation programme under the Marie
Sk\l{}odowska-Curie grant agreement no 841261.
KN acknowledges support by the European Research Council (ERC) grant agreement no 714893
and by UKRI-STFC through the University of Birmingham Particle Physics Consolidated Grant.

\bibliographystyle{JHEP}
\bibliography{bibliography}

\providecommand{\href}[2]{#2}\begingroup\raggedright\begin{thebibliography}{10}

\bibitem{spcInitial}
I.~Giomataris et~al., \emph{A novel large-volume spherical detector with
  proportional amplification read-out},
  \href{http://dx.doi.org/10.1088/1748-0221/3/09/p09007}{\emph{J. Inst.}
  {\bfseries 3} (Sep, 2008) P09007}.

\bibitem{Arnaud:2017bjh}
{\scshape NEWS-G} collaboration, \emph{{First results from the NEWS-G direct
  dark matter search experiment at the LSM}},
  \href{http://dx.doi.org/10.1016/j.astropartphys.2017.10.009}{\emph{Astropart.
  Phys.} {\bfseries 97} (2018) 54--62},
  [\href{https://arxiv.org/abs/1706.04934}{{\ttfamily 1706.04934}}].

\bibitem{Meregaglia:2017nhx}
A.~Meregaglia et~al., \emph{{Study of a spherical Xenon gas TPC for
  neutrinoless double beta detection}},
  \href{http://dx.doi.org/10.1088/1748-0221/13/01/P01009}{\emph{J. Inst.}
  {\bfseries 13} (2018) P01009},
  [\href{https://arxiv.org/abs/1710.04536}{{\ttfamily 1710.04536}}].

\bibitem{Bougamont:2015jzx}
E.~Bougamont et~al., \emph{{Neutron spectroscopy with the Spherical
  Proportional Counter based on nitrogen gas}},
  \href{http://dx.doi.org/10.1016/j.nima.2016.11.007}{\emph{Nucl. Instrum.
  Meth. A} {\bfseries 847} (2017) 10--14},
  [\href{https://arxiv.org/abs/1512.04346}{{\ttfamily 1512.04346}}].

\bibitem{GIOMATARIS2004330}
Y.~Giomataris and J.~Vergados, \emph{Neutrino properties studied with a triton
  source and a large spherical {TPC}},
  \href{http://dx.doi.org/https://doi.org/10.1016/j.nima.2004.04.223}{\emph{Nucl.
  Instrum. Meth. A} {\bfseries 530} (2004) 330 -- 358}.

\bibitem{GIOMATARIS200623}
Y.~Giomataris and J.~Vergados, \emph{A network of neutral current spherical
  {TPCs} for dedicated supernova detection},
  \href{http://dx.doi.org/https://doi.org/10.1016/j.physletb.2006.01.040}{\emph{Physics
  Letters B} {\bfseries 634} (2006) 23 -- 29}.

\bibitem{PhysRevD.79.113001}
{J. D. Vergados, F.T. Avignone III and I. Giomataris}, \emph{Coherent neutral
  current neutrino-nucleus scattering at a spallation source: A valuable
  experimental probe},
  \href{http://dx.doi.org/10.1103/PhysRevD.79.113001}{\emph{Phys. Rev. D}
  {\bfseries 79} (Jun, 2009) 113001}.

\bibitem{spcRecent}
{NEWS-G collaboration}, \emph{Spherical proportional counter: A review of
  recent developments},
  \href{http://dx.doi.org/10.1088/1742-6596/1029/1/012006}{\emph{J. Phys. Conf.
  Ser.} {\bfseries 1029} (May, 2018) 012006}.

\bibitem{sensors}
I.~Katsioulas et~al., \emph{A sparkless resistive glass correction electrode
  for the spherical proportional counter},
  \href{http://dx.doi.org/10.1088/1748-0221/13/11/p11006}{\emph{J. Inst.}
  {\bfseries 13} (Nov, 2018) P11006}.

\bibitem{Giganon:2017isb}
A.~Giganon et~al., \emph{{A multiball read-out for the spherical proportional
  counter}}, \href{http://dx.doi.org/10.1088/1748-0221/12/12/P12031}{\emph{J.
  Inst.} {\bfseries 12} (2017) P12031},
  [\href{https://arxiv.org/abs/1707.09254}{{\ttfamily 1707.09254}}].

\bibitem{ansys}
ANSYS\textregistered, ``Academic research mechanical, release 19.1.''
  \url{https://www.ansys.com/academic}.

\bibitem{Allison:2016lfl}
J.~Allison et~al., \emph{{Recent developments in {Geant4}}},
  \href{http://dx.doi.org/10.1016/j.nima.2016.06.125}{\emph{Nucl. Instrum.
  Meth. A} {\bfseries 835} (2016) 186--225}.

\bibitem{Veenhof:1998tt}
R.~Veenhof, \emph{{GARFIELD, recent developments}},
  \href{http://dx.doi.org/10.1016/S0168-9002(98)00851-1}{\emph{Nucl. Instrum.
  Meth. A} {\bfseries 419} (1998) 726--730}.

\bibitem{garfield}
H.~Schindler, ``Garfield++ user guide.''
  \url{https://garfieldpp.web.cern.ch/garfieldpp/documentation/UserGuide.pdf},
  2020.

\bibitem{heed}
I.~Smirnov, \emph{Modeling of ionization produced by fast charged particles in
  gases},
  \href{http://dx.doi.org/https://doi.org/10.1016/j.nima.2005.08.064}{\emph{Nucl.
  Instrum. Meth. A} {\bfseries 554} (2005) 474 -- 493}.

\bibitem{magboltz}
S.~Biagi, \emph{{Monte Carlo} simulation of electron drift and diffusion in
  counting gases under the influence of electric and magnetic fields},
  \href{http://dx.doi.org/https://doi.org/10.1016/S0168-9002(98)01233-9}{\emph{Nucl.
  Instrum. Meth. A} {\bfseries 421} (1999) 234 -- 240}.

\bibitem{combinedSimulation}
D.~Pfeiffer et~al., \emph{Interfacing {Geant4}, garfield++ and degrad for the
  simulation of gaseous detectors},
  \href{http://dx.doi.org/https://doi.org/10.1016/j.nima.2019.04.110}{\emph{Nucl.
  Instrum. Meth. A} {\bfseries 935} (2019) 121 -- 134}.

\bibitem{xcom}
M.~J. Berger et~al., \emph{{XCOM: Photon Cross Sections Database (version
  1.5)}},
  \href{http://dx.doi.org/https://dx.doi.org/10.18434/T48G6X}{\emph{NIST} (Mar,
  2010) }.

\bibitem{paiModel}
J.~Apostolakis et~al., \emph{An implementation of ionisation energy loss in
  very thin absorbers for the geant4 simulation package},
  \href{http://dx.doi.org/https://doi.org/10.1016/S0168-9002(00)00457-5}{\emph{Nucl.
  Instrum. Meth. A} {\bfseries 453} (2000) 597 -- 605}.

\bibitem{knoll}
G.~Knoll, \emph{Radiation Detection and Measurement}.
\newblock New York, John Wiley and Sons, Inc., 1979. 831 p., Jan, 2000.

\bibitem{shockley}
W.~Shockley, \emph{Currents to conductors induced by a moving point charge},
  \href{http://dx.doi.org/https://doi.org/10.1063/1.1710367}{\emph{J. Appl.
  Phys.} {\bfseries 9} (1938) 635--636}.

\bibitem{ramo}
S.~{Ramo}, \emph{Currents induced by electron motion},
  \href{http://dx.doi.org/10.1109/JRPROC.1939.228757}{\emph{Proceedings of the
  IRE} {\bfseries 27} (Sep, 1939) 584--585}.

\bibitem{nndc:ar37}
{National Nuclear Data Center}, ``Interactive charge of nuclides (nudat 2.8).''
  \url{https://www.nndc.bnl.gov/nudat2/}.

\bibitem{estar}
M.~J. Berger et~al., \emph{{Stopping Powers for Electrons and Positrons}},
  \href{http://dx.doi.org/10.1093/jicru/os19.2.Report37}{\emph{ICRU} {\bfseries
  os19} (Apr, 1984) }.

\bibitem{PhysRevD.98.030001}
{\scshape Particle Data Group} collaboration, \emph{Review of particle
  physics}, \href{http://dx.doi.org/10.1103/PhysRevD.98.030001}{\emph{Phys.
  Rev. D} {\bfseries 98} (Aug, 2018) 030001}.

\end{thebibliography}\endgroup
\end{document}